# Magnetic Fields and Star Formation

S Van Loo[1], T W Hartquist[2], and S A E G Falle[3]

[1] Harvard-Smithsonian Center for Astrophysics, 60 Garden Street, Cambridge, Massachusetts 02138, USA

[2] School of Physics and Astronomy, University of Leeds, Leeds LS2 9JT, UK

[3] Department of Applied Mathematics, University of Leeds, Leeds LS2 9JT, UK

Research performed in the 1950s and 1960s by Leon Mestel on the roles of magnetic fields in star formation established the framework within which he and other key figures have conducted subsequent investigations on the subject. This short tribute to Leon contains a brief summary of some, but not all, of his ground breaking contributions in the area. It also mentions of some of the relevant problems that have received attention in the last few years. The coverage is not comprehensive, and the authors have drawn on their own results more and touched more briefly on those of others than they would in a normal review. Theirs is a personal contribution to the issue honouring Leon, one of the truly great gentlemen, wits, and most insightful of astrophysicists.

Mestel and Spitzer (1956) were aware that the results of polarisation studies of the radiation from stars obscured by interstellar dust imply the presence of an interstellar magnetic field. They also cited the work of Fermi (1949) on cosmic ray acceleration in magnetised media and that of Chandrasekhar and Fermi (1953) on the equilibrium of spiral arms due to a balance of gravity by thermal, magnetic, and turbulent pressures. Mestel and Spitzer (1956) estimated the minimum mass of a cold, perfectly conducting, magnetised interstellar cloud with sufficient self-gravity to drive collapse.

Furthermore, they addressed a possibly important role, in star formation, of a non-ideal magnetohydrodynamic (MHD) mechanism called ambipolar diffusion. It is the motion, driven by electromagnetic forces, of charged particles relative to neutrals. In a weakly ionised medium the rate per unit volume at which collisions transfer momentum between neutrals and a species of charged particles is proportional to the product of the neutral number density and the number density of charged particles of that species. For a given electromagnetic force per unit volume, the charged particle drift speed relative to the neutrals increases as the inverse of the charged particle number density. This relative motion reduces the magnetic flux in the region and can result in the collapse of an object that initially had a too strong magnetic field to collapse otherwise.

In the mid-1960s, Leon wrote twin papers summarising the state of the theory of star formation (Mestel 1965a, b). In the second of these, he identified a comprehensive programme of work on the theory of star formation in a magnetised interstellar medium. One mechanism that he highlighted is the transport of angular momentum by the magnetic field, a topic that he explored further in a series of papers concluding with Mestel and Paris (1979, 1984).

Much of Leon's work on star formation concerns the evolution of a pre-formed clump that might be the immediate progenitor of a protostar. Objects called dense cores are such progenitors, and they and their evolution are addressed below. However, magnetic fields play roles in the formation of the Giant Molecular Clouds (GMCs), in which dense cores exist, and in the emergence of the dense cores themselves. Consequently, the generation of GMCs and dense cores receives attention, before the magnetic processes in dense core collapse are addressed. As a star is born and undergoes its early development, it loses mass in an outflow. The outflow from a young star drives a shock into dusty molecular material surrounding it. The theoretical and computational treatment of the shock requires the adoption of a multicomponent model in which the ambipolar diffusion of the charged dust must be included.

**Formation of Giant Molecular Clouds**

Simulations of galactic formation including feedback from star formation indicate that spiral arms can appear in the absence of magnetic fields (e. g. Agertz et al. 2011). However, the inclusion of a seed magnetic field in such a simulation leads to similar structure while amplification of the field occurs until its strength is comparable to that of the magnetic field in the diffuse phases of the interstellar medium (e. g. Wang and Abel 2009).

The Galaxy contains over a thousand Giant Molecular Clouds, which are concentrated in the spiral arms. The large-scale molecular distribution in the Rosette Molecular Cloud, a relatively nearby GMC, has been thoroughly studied (Williams et al. 1995). Its full extent is somewhat under 100 pc, and it contains of the order of $10^5$ solar masses. The mean molecular hydrogen number density, $n(H_2)$ is roughly 25 cm$^{-3}$ and most of the mass is concentrated in about 70 translucent clumps having masses ranging from a few tens to a few thousands of solar masses each and with the values of $n(H_2)$ typically about 220 cm$^{-3}$ and of T, the temperature, around 10 K. Heyer and Brunt (2012) found evidence for magnetically aligned velocity anisotropy in $^{12}$CO emission features formed in low surface brightness regions of another GMC, the Taurus Molecule Cloud, and concluded that β, the ratio of the thermal pressure to the magnetic pressure, in such regions is very small compared to unity. They did not find evidence of substantial anisotropy in $^{13}$CO emission features, which they noted probably trace regions with visual extinctions between 4 and 10 magnitudes.

Two classes of models for GMC formation exist. In the top-down models, large-scale gravitational, thermal, and magnetic instabilities in the differentially rotating disc of a galaxy trigger the origin of GMCs. The Parker instability, which leads to undulations in the magnetic field lines supporting material in a constant gravitational field perpendicular to them, is one of the instabilities that may play a key part (e. g. Mouschovias et al. 2009). In bottom-up models, compression in converging flows or compression of finite structures by supernova remnant or superbubble driven shocks trigger GMC formation. CO observations of the grand design spiral M51 favour a top-down formation mechanism (Koda et al. 2009). However, the relative importance of top-down and bottom-up mechanisms may depend on the galactic environment (e. g. Dobbs 2008).

Adopting a top-down approach, Kim and Ostriker (2002) performed MHD simulations of a self-gravitating, magnetised, differentially rotating, thin, isothermal, gaseous disc passing through a rigidly rotating gravitational potential like that of a local segment of a tightly wound trailing spiral arm. Sufficiently deep rigidly rotating potentials lead to shock formation. They examined the gravitational instability occurring under certain conditions in the postshock gas.

The Toomre Q parameter is a measure of the axisymmetric stability of a rotating unmagnetised disc of uniform density. It is given by $Q = \kappa c_s/\pi G\Sigma$ where $\kappa$, $c_s$, G, and $\Sigma$ are the epicyclic frequency, the sound speed, the gravitational constant, and the mass surface density. For a Keplerian disc, the epicyclic frequency and the orbital angular frequency, $\Omega$, are equal.

For each of several sets of parameters, e. g. spiral potential strength and preshock $\beta$, Kim and Ostriker (2002) determined the critical value of Q in the postshock region below which gravitational instability occurs. The strength of the spiral potential is parameterised by $F = 2F_1/\sin(i)$, where $F_1 = \Phi_0/(\Omega R_0)^2$, i is the pitch angle of the spiral, $\Phi_0$ is the magnitude of the depth of spiral potential, and $R_0$ is the radius of the orbit. The inclusion of the magnetic field lowers the critical value of Q. For an unmagnetised medium and $F_1 = 0$, the critical value of Q is 1, and for $\beta = 1$ and $F_1 = 0$, the critical value of Q is 0.7. For an unmagnetised medium and $F_1 = 0.06$, the critical value of Q is 2.8, and for $\beta = 1$ and $F_1 = 0.06$, the critical value of Q is 0.4. All quantities mentioned in the preceding few sentences refer to upstream, unperturbed quantities.

In addition, Kim and Ostriker (2002) showed that the nonlinear growth of self-gravitating perturbations results in the emergence of spur-like structures in the magnetised models. Such structures are commonly observed in spiral galaxies. Kim and Ostriker (2002) attributed the growth of the spurs to the magneto-Jeans mechanism, which leads to the magnetic tension force opposing the Coriolis force. In the absence of a magnetic field, the Coriolis force would prevent the coalescence of matter along a spiral arm. Consequently, magnetic effects seem essential for spur formation. The spurs in the magnetised models undergo fragmentation to form $4 \times 10^6$ solar mass clumps, which may fragment further into GMCs.

In attempts to understand how GMCs might form in bottom-up scenarios, several groups have simulated the production of GMC-like cloud material in supersonic collisions of streams of magnetised gas that is initially in the warm phase with temperatures of several thousand K. Heitsch et al. (2009) conducted one of the more recent studies of this type. Vázquez-Semadeni et al. (2011) have included ambipolar diffusion in such simulations. For a given heating rate per particle, the warm phase exists only at pressures below a specific maximum pressure determined by that heating rate, and at pressures above that specific maximum pressure the cloud phase is the only stable phase. In the local interstellar medium the thermal pressure of the warm gas and the atomic clouds is only a factor of a few or less below the relevant maximum pressure, which is about a few eV cm$^{-3}$. A moderately supersonic collision of two streams of the warm gas would lead to most of it undergoing a phase transition and becoming cold. Shocks exist on either side of the contact discontinuity separating the material in the two streams. In a non-magnetic simulation, the material in one stream passes through one of these shocks, while the material in the other stream passes through the other. The shocked gas is at a higher pressure than the unshocked gas in the streams. It becomes thermally unstable and cools. If the relative velocity of the unshocked material in the two streams is along a large-scale uniform magnetic field, the cooling leads to a compression of about a hundred or more. If the preshock $\beta$ has a value of unity and the relative velocity of the unshocked material in the streams is perpendicular to a large-scale magnetic field, the compression is more modest and the postshock regions have low values of $\beta$, which depend on the Mach number of the relative velocity.

The roles of slow-mode shocks and fast-mode shocks in bottom-up models of GMC formation were revealed in axisymmetric simulations, performed by Van Loo et al. (2007), of shocks interacting with individual initially spherical uniform clouds. Three wave modes exist in ideal MHD. Unsurprising, the fast-mode waves are the fastest. A different type of shock is associated with each of the wave-modes, and the minimum speed differs for each wave-mode and depends on the angle between the wavevector and the upstream magnetic field. Note that intermediate-mode shocks are not evolutionary (e. g. Falle and Komissarov 2001) and should be ignored.

In each of the Van Loo et al. (2007) simulations, a shock initially propagates into a tenuous upstream medium in pressure equilibrium with a spherical cloud of warm phase gas. The upstream magnetic field is parallel to the shock velocity and is uniform in the upstream tenuous medium and the spherical cloud. Compression during cooling behind the fast-mode shock is moderate, and the cooled postshock material is at low $\beta$, as is consistent with the results of the observational study of the Taurus Molecular Cloud mentioned above. The compression behind the slow-mode shock, which follows, is substantial, and behind it $\beta$ approaches unity. The general morphology of the simulated shocked cloud is similar to that shown in a CO map of the W3 molecular cloud.

Van Loo et al. (2010) have produced three-dimensional simulations that are analogous to those of Van Loo et al. (2007). One conclusion drawn by Van Loo et al. (2010) is that a shock propagating moderately obliquely to the magnetic field produces a cooled cloud similar to a cloud formed behind a shock that propagates perpendicular to the magnetic field. Another significant conclusion is that the fast-mode Mach number must be between roughly two to three, if the cold cloud is to be reasonably long-lived and have a substantial fraction of its volume contained in regions of low $\beta$. Shocks with higher Mach numbers would lead to cold cloud lifetimes that are too short. Weaker shocks do not result in much of the cloud material having a particularly small $\beta$.

**Formation of Dense Cores**

Dense cores, having temperatures of about 10 K and masses of roughly 1 to 10 solar masses each and in which n(H$_2$) is roughly $10^4$ to $10^5$ cm$^{-3}$ (e. g. Benson and

Myers 1989), are considered to be the progenitors of low-mass stars: those stars with less than four solar masses. More massive cores in which clusters of stars are born also exist. Of course, self-gravity is important in the formation of clusters. Fiedler and Mouschovias (1993) included ambipolar diffusion in an axisymmetric model of the formation of a flattened, magnetised dense core due to the gravitational collapse of a more diffuse distribution of gas. They also followed the evolution until the central number density was several orders of magnitude higher than those mentioned above.

There exists a vast literature on the formation, due to the evolution of nonlinear MHD perturbations, of structures within clumps having properties similar to the translucent clumps in GMCs (e. g. Ballesteros-Paredes et al. 2007). To understand what processes occur in such simulations, it is normally useful to perform a linear analysis, similar that conducted by Mouschovias et al. (2011), to gain insight into waves in weakly ionised media. Many of the ideal MHD simulations of the formation of core-like structures have been based on the assumption that the perturbations are on background media in which $\beta$ is small. Thus, Falle and Hartquist (2002) performed a linear analysis of ideal MHD perturbations in a medium with a small $\beta$. If $\beta$ is small, in a fast-mode, linear perturbation, the ratio of the magnitude of the density perturbation to the density is of the same order as the ratio of the magnitude of the velocity perturbation to the Alfvén speed. In a slow-mode, linear perturbation in a low $\beta$ region, the ratio of the magnitude of the density perturbation to the density is of the same order of $\beta^{-1/2}$ times the ratio of the magnitude of the velocity perturbation to the Alfvén speed. Falle and Hartquist (2002) reported the results of plane-parallel, time-dependent simulations showing that the nonlinear steepening of a fast-mode wave with a finite but modest amplitude can readily excite slow-mode waves as long as the angle between the fast-mode wavevector and the magnetic field is neither too large nor too small. The slow-mode excitation produces persistent inhomogeneities with large density contrasts.

In an extension of the work of Falle and Hartquist (2002), Lim et al. (2005) included ambipolar diffusion and took each initial perturbation to be a nonlinear fast-mode wave. The neutral-ion momentum transfer timescale is the timescale over which a neutral particle, subjected to no forces other than friction due to collisions with ions, decelerates. It is inversely proportional to the ion number density. Lim et al. (2005) defined a lengthscale, $l_d$, given by that timescale times the speed of Alfvén waves having frequencies that are much lower than the inverse of that timescale. In a translucent clump or dense core the value of $l_d$ would typically be of order 0.01 pc. Lim et al. (2005) found that in a medium with $\beta$ of about 0.01, ambipolar diffusion substantially reduces the maximum density arising from the evolution of any fast-mode wave with an initial wavelength of less than several hundred times $l_d$ and an initial velocity amplitude having a magnitude less than the Alfvén speed.

Van Loo et al. (2008) carried out two-dimensional simulations to examine the effect of ambipolar diffusion on the formation of structure through the generation of slow-mode disturbances by the non-linear steepening of initial fast-mode perturbations. A number of their results are in harmony with those of Lim et al. (2005). In addition, they argued that substructure within cores, like that seen in core D of TMC-1 (Peng et al. 1998), can be generated by slow-mode excitation due to the nonlinear steepening of fast-mode waves only in cores with unusually high fractional ionisations. They also noted that the presence of fast-mode perturbations in the intercore gas and the association of dense cores with slow-mode structures account for the transition from superthermal linewidths in the intercore medium to thermal linewidths in the cores (Myers 1983).

Li et al. (2012) have incorporated ambipolar diffusion in three-dimensional simulations of the power spectrum of MHD waves in a background medium with $\beta$ = 0.1. The maximum amplitudes are sub-Alfvénic but superthermal. Using ideal MHD results and non-ideal MHD results as input, Li et al. (2012) have performed radiative transfer calculations to produce synthetic CS J = 2–1 and $H^{13}CO^+$ J = 1–0 line profiles. They have concluded that an observable narrowness of ion line profiles relative to neutral line profiles can arise due to ambipolar diffusion. Downes (2012) has also addressed the effects of ambipolar diffusion on the wave spectrum in a star forming region and has included dust grains as a separate fluid.

Magnetic reconnection in a medium with a spectrum of nonlinear waves, like that studied by Li et al. (2012), has received attention recently (e. g. Santos-Lima et al. 2010). However, Santos-Lima et al. (2010) have considered Ohmic dissipation only and have neglected ambipolar diffusion. This is an important distinction since Ohmic diffusiion can lead to reconnection, whereas ambipolar diffusion cannot .

**Dense Core Collapse and Angular Momentum Loss**

The self-gravity of a dense core plays a central part in star formation. An isothermal, spherical, uniform density, perfectly conducting core threaded by a uniform magnetic field and surrounded by a very tenuous medium having a constant pressure is not in equilibrium. Mouschovias (1976) found isothermal, perfectly conducting, axisymmetric equilibrium states with the same distributions of mass on the magnetic field lines as the corresponding uniformly magnetised spheres. Mouschovias and Spitzer (1976) showed that such an axisymmetric core with a mass-to-magnetic flux ratio below a critical value does not collapse due to self-gravity, even if it is cooled to absolute zero and the surrounding tenuous medium has an indefinitely high uniform pressure. That critical value is $0.53(5/G)^{1/2}/3\pi$. Structures with a mass-to-flux ratio above (below) that are said to be magnetically supercritical (subcritical). Mouschovias and Tassis (2010) have addressed the difficulty in inferring from observational data whether a core is magnetically supercritical or subcritical.

If embedded in surroundings with sufficiently high pressures, a supercritical core will collapse on roughly a gravitational free-fall time. In such a case, the main role of the magnetic field will be in angular momentum transport. Differential rotation will drive MHD waves that carry away angular moment (Mestel 1965b; Mestel and Paris 1979; 1984).

The collapse of a magnetically subcritical core, as ambipolar diffusion leads to an increase in the mass-to-flux ratio, has received considerable theoretical attention. Basu and Mouschovias (1995 a, b) presented results of axisymmetric thin disc simulations of ambipolar-diffusion regulated collapse including angular momentum transport for a wide range of parameters. The vertical structure of a disc was assumed to be in static equilibrium. They adopted a two-fluid description. At the same time Ciolek and Mouschovias (1995) published results of multi-fluid, axisymmetric simulations of the ambipolar-diffusion regulated collapse of non-rotating thin discs. They included neutral-gas, ion, electron, charged-grain, and neutral-grain fluids and adopted a chemical network in order to calculate the ion and electron abundances and the charge distributions on grains. At fractional ionisations of about $10^{-8}$ and below, the frequency at which a neutral collides with grains is greater than that at which it collides with ions; then grain-neutral drag dominates over ion-neutral drag and controls the ambipolar diffusion. An additional complication arises because the timescale in a dense core for a 100 nm grain to encounter a mass of neutral particles comparable to its own is about the same as its gyroperiod. The inverse of the ratio of those two timescales is called the Hall parameter.

In the general picture that has emerged, ambipolar-diffusion induced dynamics results in an outwardly propagating C-type MHD shock (Li and McKee 1996), which is sometimes referred to as the "magnetic diffusion shock". In axisymmetric models neutral material, closer to the symmetry axis than this shock, falls inwardly at nearly the free-fall speed until the centrifugal force becomes important and triggers the formation of a hydrodynamic shock that decelerates the infalling matter and potentially allows a Keplerian disc to form (Shu et al. 1987).

To show that the angular momentum transport during the formation of a low-mass star is neither too efficient nor too inefficient for a protoplanetary disc to develop, Dapp et al. (2012) have recently extended the work of Basu and Mouschovias (1995a, b) by including the effects of neutral and positively and negatively charged grains on ambipolar diffusion. Rather than computationally solve fluid equations for the ions, electrons, and grains, they derived appropriate diffusion terms to incorporate in the magnetic induction equation. Such terms moderate the effectiveness of the angular momentum transfer. The calculation of those terms required that the inclusion of a network for the chemistry involving charged species. This approach allowed them to study a problem in which the highest density is sixteen orders of magnitude higher than the lowest density. They found that ambipolar diffusion dominates diffusion up to a number density of about $5 \times 10^{12}$ cm$^{-3}$, but at higher densities Ohmic dissipation is more important. Their primary result is the demonstration that protoplanetary discs can form in a model incorporating ambipolar diffusion and Ohmic dissipation.

Dapp et al. (2012) did not include the Hall diffusion term in their magnetic induction equation. The Hall term is associated with a current component that, in the neutral frame, is perpendicular to the magnetic field and the electric field. It is important when the Hall parameter is of order unity or smaller. Braiding and

Wardle (2012) have found similarity solutions describing the evolution of a magnetised, rotating, thin, isothermal disc in an analysis including Hall diffusion and diffusion due to a combination of ambipolar diffusion and Ohmic dissipation. Braiding and Wardle (2012) have concluded that reasonable assumptions about the Hall diffusion can lead to a change of the size of the protoplanetary disc that appears by up to an order of magnitude relative to what it would be if Hall diffusion were not included.

Another source of diffusivity that might affect angular momentum transport has been considered by Santos-Lima et al. (2012), They have suggested that magnetic reconnection, occurring due to the presence of a spectrum of nonlinear waves, can result in an effective diffusivity that would reduce the efficiency of the angular momentum transport by MHD waves arising due to differential rotation. Seifried et al. (2012b) have argued that reconnection is unimportant but that the presence of a background spectrum of nonlinear waves affects the efficiency of angular momentum transportation by MHD waves in any case.

While Dapp et al. (2012) and Braiding and Wardle (2012) adopted a thin disc approximation, Duffin and Pudritz (2009) studied protoplanetary disc formation by performing a three-dimensional, Adaptive Mesh Refinement, MHD simulation including ambipolar diffusion. The initial state was magnetically supercritical and similar to a rotating Bonner-Ebert sphere, and Duffin and Pudritz (2009) did not take grains into account when calculating the ambipolar diffusivity. They appear to have neglected the Ohmic and Hall diffusivities. They found that a disc formed and that a two-sided jet also developed. They also considered the formation of binaries and drew conclusions similar to those of Hosking and Whitworth (2004). Those authors' three-dimensional, MHD simulation, including ambipolar diffusion, of the collapse of an initially magnetically subcritical core had already indicated that the magnetic removal of angular momentum inhibits fragmentation. Ideal MHD simulations performed by Price and Bate (2007) also imply that the transport of angular momentum by magnetic fields suppresses fragmentation, but they concluded that fragments appear if sufficiently large initial perturbations are present.

**Outflows, Multifluid Magnetohydrodynamic Shocks in Dusty Media, and Ionisation Fronts**

As mentioned above, Duffin and Pudritz (2009) found that a jet developed in their simulations. Seifried et al. (2012a) have worked on the development of a general outflow criterion with the application to the early phases of high-mass star formation, particularly in mind. They performed simulations with different initial rotational and magnetic energies. Initially weak fields or high rotational energies lead to well-collimated, fast jets. In contrast, initially strong fields result in poorly collimated, low-velocity outflows. Fast jets are associated with Keplerian protostellar discs. All of the outflows are launched from the discs by centrifugal acceleration, and the toroidal magnetic field component contributes increasingly to the gas acceleration further away from the discs. The poor collimation of the outflows in runs with strong initial magnetic fields is due to the hoop stresses being weak as a consequence of the slow build-up of a toroidal magnetic field component when the disc rotation is markedly sub-Keplerian.

Outflows are ubiquitous in star forming regions and drive shocks into the magnetised, weakly ionised media surrounding young stars. Mullan (1971) recognised the importance of ambipolar diffusion in shocks driven into weakly ionised astrophysical media. Draine (1980) and Chernoff et al. (1982) included a great deal of the relevant microphysics into shock models incorporating ambipolar diffusion. The compilation of equations and rates by Draine et al. (1983) enabled others to develop multifluid MHD models of shocks in star forming regions.

In a weakly ionised, magnetised star forming region, a shock with a speed of less than about 40 km s$^{-1}$ contains a smooth flow in which none of the ion, electron, neutral, or grain components develop thin subshocks with thicknesses established by scattering mean free paths. Such shocks with continuous flow variables are called C-type shocks. They are important for feedback in star forming regions, and sputtering and grain-grain collisions in them inject material contained in grain mantles and refractory cores into the gas phase (e. g. Schilke et al. 1997; Gusdorf et al. 2008).

The papers cited in the previous two paragraphs all concern models of perpendicular shocks, those that propagate perpendicular to the upstream magnetic field. Though the treatment of grain dynamics used by Draine et al. (1983) and authors who followed their approach is adequate for a wide range of interesting shock conditions, Pilipp et al. (1990) demonstrated the necessity of treating grain dynamics in perpendicular shock models more rigorously in many situations in which the Hall parameter of the grains is of order unity or less. They treated each grain component as a fluid. The most sophisticated perpendicular shock models are those used by Guillet et al. (2011) to study grain destruction. They employed a Monte Carlo approach to follow the grain dynamics and charging, and included all components of the currents carried by the grains in the calculation of the magnetic field.

Pilipp and Hartquist (1994) recognised the necessity for the development of a treatment of oblique shocks comparable to that introduced by Pilipp et al. (1990) for perpendicular shocks. Wardle (1998) elucidated the grounds for difficulties encountered by Pilipp and Hartquist (1994) when trying to obtain solutions corresponding to fast-mode shocks. Those grounds presented challenges for the construction of fast-mode shock models in situations when local equilibrium does not obtain. Falle (2003) developed a technique, suitable for appropriate values of the grain Hall parameters, for dealing with the model equations governing the time-dependent evolution of a plane-parallel oblique MHD shock propagating into a weakly ionised dusty medium. Van Loo et al. (2009) took the first step to apply it with the inclusion of a treatment of the grain charging, ionisation structure, and cooling. Ashmore et al. (2010) and Van Loo et al. (2012) used the technique in the study of the interactions of oblique shocks with density variations and in the investigation of the dependence of grain destruction on the obliqueness of the shock, respectively.

Chen and Ostriker (2012) have argued that as a C-type shock forms, for example in a collision between two clumps or cores, the mass-to-flux ratio increases throughout some of the shock structure. The increase will persist for a time comparable to the final steady C-shock thickness divided by the drift speed, through the shock, of the ions with respect to the neutrals. They have suggested that under favorable conditions, the transient regions may be magnetically supercritical and collapse on timescales shorter than the relevant timescales for the shocks to evolve enough for the mass-to-flux ratios to decrease substantially. Thus, star formation would be triggered.

High mass stars affect their environments with their radiation fields as well as their outflows. Williams and Dyson (2001) examined the internal structures of stationary ionisation fronts with oblique upstream magnetic fields and found solutions only for upstream parameters in a range allowed by the evolutionary conditions that they identified. Overheating in some fronts can lead to some of the solutions not corresponding to resolved internal structures, even if those solutions are allowed by the evolutionary conditions. The inclusion of subshocks can lead to some, but not all, of these jumps being realised.

A number of three-dimensional simulations of magnetised ionisation fronts have been reported. Krumholz et al. (2007) performed the first three-dimensional, global simulation of an H II region expanding into a magnetised medium and found that magnetic fields suppress the sweeping up of gas perpendicular to magnetic field lines. This leads to small density contrasts and weak shocks at the leading edge of an H II region's expanding shell. Henney et al. (2009) reported the results of three-dimensional MHD simulations of the photoionisation of a molecular globule by an external ultraviolet source. They showed that, for a strong ionising field, significant deviations from the non-magnetic evolution are seen if the pressure of the initial magnetic field threading the globule is greater than a hundred times the gas pressure. Mackey and Lim (2011) investigated the effects of initially uniform magnetic fields on the formation and evolution of dense pillars and cometary globules at the boundaries of H II regions. Like Henney et al. (2009), they found that only a strong initial magnetic field can significantly alter the dynamics. From simulations of a magnetised, collapsing region in which radiative feedback occurs, Price and Bate (2009) concluded that a strong magnetic field and radiative feedback leads to a star formation rate of less than or about equal to ten percent per free-fall time.

## Conclusions

This brief review contains multiple references to work based on numerical computations that would have been unfeasible when Leon was doing his earliest seminal work on magnetic fields in star formation. That the purposes of many of the recent calculations have been the exploration of mechanisms considered by Leon is a testament to his physical insight.

## Acknowledgements

The authors are grateful to the UK Science and Technology Funding Council for support. SVL greatly appreciates funding provided by the Smithsonian SMA Fellowship he currently holds.